\documentclass[12pt,a4paper]{article}

\usepackage{latexsym}
\usepackage{amsfonts}
\usepackage{amssymb}
\usepackage{amsmath}
\usepackage[mathscr]{eucal}
\usepackage{theorem}
\usepackage{enumerate}
\usepackage{cite}
\usepackage{url}
\usepackage[dvips]{color}
\usepackage{graphicx}
\usepackage{xypic}

\theoremstyle{plain}
\theorembodyfont{\itshape}
\newtheorem{thm}{Theorem}

\newtheorem{prp}{Proposition}

\newtheorem{dfn}{Definition}

\theorembodyfont{\upshape}
\newtheorem{exe}{Example}

\newcommand{\proof}{\noindent {\bf Proof:} \hspace{0.1in}}
\newcommand{\qed}{\hfill\mbox{\raggedright $\Box$}\medskip}

\newcommand{\mydate}{
 \ifcase\month \or
 January\or February\or March\or April\or May\or June\or
 July\or August\or September\or October\or November\or December\fi
 \space \number\year}
\newcommand{\smin}{\,\raisebox{0.06em}{${\scriptstyle \in}$}\,}

\newcommand{\ssmin}{\,\raisebox{0.06em}{${\scriptscriptstyle \in}$}\,}

\newcommand{\bwedge}{\raisebox{0.2ex}{${\textstyle \bigwedge}$}}
\newcommand{\bvee}{\raisebox{0.2ex}{${\textstyle \bigvee}$}}

\newcommand{\mathsl}[1]{\mbox{\textsl{\textsf{#1}}}}

\newcommand{\grpd}[3]{\ensuremath{%
 \xymatrix@1{{#2} \; & \ar[l]_{~#3} \; {#1}}}}

\begin{document}

\title{Local Symmetries in Gauge Theories \\
       in a Finite-Dimensional Setting}
\author{Michael Forger\,\thanks{E-mail: \textsf{forger@ime.usp.br}}~~~and~
        Bruno Learth Soares\,\thanks{E-mail: \textsf{bsoares@ime.usp.br}}}
\date{\normalsize
      Departamento de Matem\'atica Aplicada, \\
      Instituto de Matem\'atica e Estat\'{\i}stica, \\
      Universidade de S\~ao Paulo, \\
      Caixa Postal 66281, \\
      BR--05315-970~ S\~ao Paulo, S.P., Brazil
}
\maketitle

\thispagestyle{empty}

\begin{abstract}
\noindent
 It is shown that the correct mathematical implementation of symmetry in
 the geo\-metric formulation of classical field theory leads naturally
 beyond the concept of Lie groups and their actions on manifolds, out
 into the realm of Lie group bundles and, more generally, of Lie
 groupoids and their actions on fiber bundles. This applies not only
 to local symmetries, which lie at the heart of gauge theories, but
 is already true even for global symmetries when one allows for fields
 that are sections of bundles with (possibly) non-trivial topology or,
 even when these are topologically trivial, in the absence of a preferred
 trivialization.
\end{abstract}

\begin{flushright}
 \parbox{12em}{
  \begin{center}
   Universidade de S\~ao Paulo \\
   RT-MAP-0901 \\
   \mydate
  \end{center}
 }
\end{flushright}

\newpage

\setcounter{page}{1}

\section{Introduction}

Gauge theories constitute a class of models of central importance in
field theory since they provide the conceptual basis for our present
understanding of three of the four fundamental interactions~-- strong,
weak and electromagnetic. At the very heart of gauge theories lies the
principle of gauge invariance, according to which physics is invariant
under symmetry transformations even if one is allowed to perform
different symmetry transformations at different points of space-time:
such transformations have come to be known as local symmetries, as
opposed to rigid transformations which are the same at all points
of space-time and are commonly referred to as global symmetries.%
\footnote{In this paper, when speaking of symmetries (local or global),
we are tacitly assuming that we are dealing with continuous symmetries,
not with discrete ones.}

One of the reasons why gauge theories are so natural is that there is a
standard procedure, due to Hermann Weyl, for ``gauging'' a global symmetry
so as to promote it to a local symmetry, or to put it differently, for
constructing a field theory with local \linebreak symmetries out of any
given field theory with global symmetries. A~salient feature of this
method is that it requires the introduction of a new field, the gauge
potential, which is needed to define covariant derivatives that replace
ordinary (partial) derivatives: such a prescription, known as ``minimal
coupling'', is already familiar from general relativity. \linebreak
(A subsequent step is to provide the gauge potential with a dynamics
of its own.) \linebreak In his original proposal~\cite{We1}, Weyl
explored the possibility to apply this construction to scale trans%
formations and, by converting scale invariance into a local symmetry,
arrive at a unified theory of gravity and electromagnetism.
Although this version was almost immediately dismissed\,%
\footnote{Somewhat ironically, an important remnant of this very first
attempt at a unification between the fundamental interactions (the only
two known ones at that time) is the persistent use of the word ``gauge''.}
after Einstein had argued that it leads to physically unacceptable
predictions, the method as such persisted. It became fruitful after the
advent of quantum mechanics, when in his modified proposal~\cite{We2},
Weyl applied the same construction to phase transformations and showed
that converting phase invariance, which is a characteristic feature
of quantum mechanics, into a local symmetry, the electromagnetic
field (or better, the electromagnetic potential) emerges naturally.
In this way, Weyl created the concept of a gauge theory and established
electromagnetism (coupled to matter) as its first example. \linebreak
In the 1950's, these ideas were extended from the abelian group $U(1)$
of quantum mechanical phases to the nonabelian isospin group $SU(2)$~%
\cite{YM} and, soon after, to general compact connected Lie groups~%
$G$~\cite{Ut}.

Another aspect that deserves to be mentioned in this context is that the
field theory governing the only one of the four fundamental interactions
not covered by gauge theories of the standard Yang-Mills type, namely
Einstein's general relativity, also exhibits a kind of local symmetry
(even though of a slightly different type), namely general coordinate
invariance. The same type of local symmetry, going under the name of
reparametrization invariance, prevails in string theory and membrane
theory. Thus we may say that the concept of local symmetry pervades
all of fundamental physics.

Unfortunately, there is one basic mathematical aspect of local symmetries
which is the source of innumerous difficulties: the relevant symmetry
groups are infinite-dimensional. For example, on an arbitrary space-time
manifold~$M$, gauging a field theory which is invariant under the action of
some connected compact Lie group~$G$ will lead to a field theory which, in
the simplest case where all fiber bundles involved are globally trivial, is
invariant under the action of the infinite-dimensional group $C^\infty(M,G)$.
Similarly, in general relativity, we find invariance under the action of the
infinite-dimensional group $\mathrm{Diff}(M)$~-- the diffeomorphism group
of space-time~\cite{HE}. The same type of local symmetry group also appears
in string theory and membrane theory, although in this case $M$ is to be
interpreted as the parametrizing manifold and not as space-time.
As is well known, the mathematical difficulties one has to face
when dealing with such infinite-dimensional groups and their actions
on infinite-dimensional spaces of field configurations or of solutions
to the equations of motion (covariant phase space) are enormous and
often insurmountable, in particular when $M$ is not compact, as is
the case for physically realistic models of space-time~\cite{HE}.

In view of this situation, it would be highly desirable to recast the
property of invariance of a field theory under local symmetries into
a form where one deals exclusively with finite-dimensional objects.
That such a reformulation might be possible is suggested by observing
that gauge transformations are, in a very specific sense, localized
in space-time: according to the principle of relativistic causality,
performing a gauge transformation in a certain region can have no
effect in other, causally disjoint regions. Intuitively speaking,
gauge transformations are ``spread out'' over space-time, and this
should make it possible to eliminate all reference to infinite-%
dimensional objects if one looks at what happens at each point of
space-time separately and only fits the results together at the very end.

This idea can be readily implemented in mechanics, where space-time~$M$ is
reduced to a copy of the real line $\mathbb{R}$ representing the time axis.
In the context of the lagrangian formulation, the procedure works as follows.
Consider an autonomous mechanical system\,%
\footnote{The generalization to non-autonomous systems is straightforward.}
with configuration space~$Q$ and lagrangian~$L$, which is a given function
on the tangent bundle $TQ$ of~$Q$: its dynamics is specified by postulating
the solutions of the equations of motion of the system to be the stationary
points of the action functional~$S$ associated with an arbitrary time
interval $[t_0,t_1]$, defined by
\begin{equation} \label{eq:ACTMEC}
 S[\mathsl{q}]~=~\int_{t_0}^{t_1} dt~L(\mathsl{q}(t),\dot{\mathsl{q}}(t))
\end{equation}
for curves $\; \mathsl{q} \smin C^\infty(\mathbb{R},Q) \;$ in~$Q$. 
To implement the notion of symmetry, we must fix a Lie group~$G$
together with an action
\begin{equation} \label{eq:ACTGQ}
 \begin{array}{ccc}
  G \times Q & \longrightarrow &     Q     \\[1mm]
     (g,q)   &   \longmapsto   & g \cdot q
 \end{array}
\end{equation}
of~$G$ on~$Q$ and note that this induces an action
\begin{equation} \label{eq:ACTGTQ}
 \begin{array}{ccc}
    G \times TQ   & \longrightarrow &
                      TQ                            \\[1mm]
  (g,(q,\dot{q})) &   \longmapsto   &
  g \cdot (q,\dot{q}) = (g \cdot q,g \cdot \dot{q}) 
 \end{array}
\end{equation}
of~$G$ on the tangent bundle~$TQ$ of~$Q$ as well as, more generally, an action
\begin{equation} \label{eq:ACTTGTQ}
 \begin{array}{ccc}
      TG \times TQ     & \longrightarrow &
                      TQ                            \\[1mm]
  ((g,Xg),(q,\dot{q})) &   \longmapsto   &
  (g,Xg) \cdot (q,\dot{q}) = (g \cdot q \,,\, g \cdot \dot{q} + X_Q(g \cdot q))
 \end{array}
\end{equation}
of the tangent group~$TG$ of~$G$ on the tangent bundle~$TQ$ of~$Q$. (Here,
we use that the tangent bundle $TG$ of a Lie group~$G$ is again a Lie group,
whose group multiplication is simply the tangent map to the original one,
and that we can use, e.g., right translations to establish a global
trivialization
\begin{equation} \label{eq:TRITG}
 \begin{array}{ccc}
       TG     & \longrightarrow & G \times \mathfrak{g} \\[1mm]
  (g,\dot{g}) &   \longmapsto   &  (g,\dot{g} g^{-1})
 \end{array}~,
\end{equation}
which shows that $TG$ is isomorphic to the semidirect product of~$G$ with
its own Lie algebra $\mathfrak{g}$ and allows to bring the induced action
of $TG$ on $TQ$, which is simply the tangent map to the original action
of~$G$ on~$Q$, into the form given in eqn~(\ref{eq:ACTTGTQ}).)
Then the system will exhibit a global symmetry under~$G$ if~$S$ is
invariant under the induced action of~$G$ on curves in~$Q$, that is,
if $\; S[g \cdot \mathsl{q}] = S[\mathsl{q}] \;$  where $\; (g \cdot
\mathsl{q})(t) = g \cdot \mathsl{q}(t)$; obviously, this will be
the case if and only if the lagrangian~$L$ is invariant under the
action~(\ref{eq:ACTGTQ}) of~$G$ on~$TQ$. On the other hand, the
system will exhibit a local symmetry under~$G$ if~$S$ is invariant
under the induced action of curves in~$G$ on curves in~$Q$, that
is, if $\; S[\mathsl{g} \cdot \mathsl{q}] = S[\mathsl{q}] \;$ where
$\; (\mathsl{g} \cdot \mathsl{q})(t) = \mathsl{g}(t) \cdot \mathsl{q}(t)$.
\linebreak
Now it is easily verified that this will be so if and only if the
lagrangian~$L$ is invariant under the action~(\ref{eq:ACTTGTQ}) of~%
$TG$ on~$TQ$ (see, e.g., Ref.~\cite{FK}). In other words, the condition
of invariance of the action  functional under the infinite-dimensional
group $C^\infty(\mathbb{R},G)$ can be reformulated as a condition of
invariance of the lagrangian under a finite-dimensional Lie group,
which is simply the tangent group~$TG$ of the original global
symmetry group~$G$. Moreover, it is also shown in Ref.~\cite{FK}
how one can use this approach to ``gauge'' a given global symmetry to
promote it to a local symmetry, provided one replaces the configuration
space~$Q$ by its cartesian product with the Lie algebra $\mathfrak{g}$
and studies the dynamics of curves $(\mathsl{q},\mathsl{A})$ where
$\mathsl{A}$ is a Lagrange multiplier: the mechanical analogue of
the gauge potential. Of course, in mechanics there is
no natural dynamics for such a Lagrange multiplier, since the
``curvature'' of this ``connection'' vanishes identically.

The main goal of the present paper, which is based on the PhD thesis
of the second author~\cite{So}, is to show how one can implement
the same program~-- recasting local symmetries in a purely finite-%
dimensional setting~-- in field theory, that is, for full-fledged
gauge theories and in a completely geometric setup. As we shall see,
this requires an important extension of the mathematical tools
employed to describe symmetries: the passage from Lie groups and
their actions on manifolds to Lie group bundles and their actions
on fiber bundles (over the same base manifold).

The formulation of (classical) gauge theories in the language of
modern differential geo\-metry is an extensive subject that, since
its beginnings in the 1970's, has been addressed by many authors;
some references in this direction which have been useful in the
course of our work are~\cite{DV,Bl,Na,Ga,MM,Be}. It should also
be mentioned that Lie group bundles can be regarded as a special
class of Lie groupoids, namely, locally trivial Lie groupoids for
which the source projection and the target projection coincide~%
\cite{MK}, and the use of Lie groupoids in gauge theories has
been advocated before by some authors~\cite{Me,Ke}. However,
we believe that the theory of Lie group bundles should not be
regarded as just a special case of the theory of general Lie
groupoids but has a flavor of its own. In particular, there
are various constructions which are familiar from the theory
of Lie groups that can be extended in a fairly straightforward
manner to Lie group bundles but are far more difficult to
formulate for general Lie groupoids. As an example which is
relevant here, consider the fact used above that the tangent
bundle $TG$ of a Lie group~$G$ is a Lie group and the tangent
map to an action of a Lie group~$G$ is an action of the tangent
Lie group~$TG$: its natural extension to Lie group bundles is
formulated in Theorems~\ref{thm:LGBJ1} and~\ref{thm:LGBJ2} below,
but the question of how to extend it to general Lie groupoids~--
i.e., how to define jet groupoids of Lie groupoids and their
actions~-- does not seem to have been addressed anywhere in
the literature. Another example would be the question of how
to define the notion of invariance of a tensor field under
the action of a Lie groupoid. And finally, similar statements
hold concerning the question of applications: there are lots
of lagrangians in physics that are gauge invariant, but none
that are gauge plus space-time diffeomorphism invariant.
True Lie groupoids may become relevant if one wants to
unify internal symmetries with space-time symmetries,
but strong restrictions will have to be imposed on the
space-time part: the question of how to formulate this
as elegantly and concisely as possible is presently under
investigation. At any rate, we think that, intuitively
speaking, Lie group bundles occupy a place ``half way in
between'' Lie groups and general Lie groupoids and that
they deserve a separate treatment and a place in their
own right.

\section{Jet bundles and the connection bundle}

The theory of jet bundles~-- as exposed, e.g., in Ref.~\cite{Sa}~--
is an important tool in differential geometry and, in particular,
plays a central role for the understanding of symmetries in gauge
theories as advocated in this paper. The general definition of
jets (as equivalence classes of local maps whose Taylor expansions
coincide up to a certain order) is somewhat complicated but will
not be needed here since in first and second order (which is all
we are going to consider) there are alternative definitions that
are simpler~\cite{FR}. In fact, given a fiber bundle $E$ over a
manifold $M$ with projection $\, \pi: E \longrightarrow M$, we
can define its first order jet bundle $JE$ and its linearized
first order jet bundle $\vec{J} E$ as follows: for any point
$e$ in~$E$ with base point $\, m = \pi(e) \,$ in~$M$, consider
the affine space
\begin{equation} \label{eq:FJB1}
 J_e E~=~\{ \, u \smin L(T_m^{} M,T_e^{} E) \, | \;
              T_e^{} \pi \circ u~=~\mathrm{id}_{T_m^{} M}^{} \, \}~,
\end{equation}
and its difference vector space
\begin{equation} \label{eq:LJB1}
 \vec{J}_e E~=~L(T_m^{} M,V_e^{} E)~=~T_m^* M \otimes V_e^{} E~,
\end{equation}
where $\, V_e^{} E = \ker T_e^{} \pi \,$ is the vertical space of~$E$
at~$e$, and note that this defines $JE$ as an affine bundle over~$E$
and $\vec{J} E$ as a vector bundle over~$E$ with respect to the target
projection (which takes $J_e E$ and $\vec{J}_e E$ to $e$) but also
defines both of them as fiber bundles over~$M$ with respect to the
source projection (which takes $J_e E$ and $\vec{J}_e E$ to $\, m
= \pi(e)$). Moreover, composition with the appropriate tangent maps
provides a canonical procedure for associating with every strict
homomorphism $\, f: E \longrightarrow F \,$ of fiber bundles $E$
and $F$ over~$M$ a homomorphism $\, Jf: JE \longrightarrow JF \,$
of affine bundles (sometimes called the jet prolongation of~$f$) and
a homomorphism $\, \vec{J} f: \vec{J} E \longrightarrow \vec{J} F \,$
of vector bundles covering~$f$: in this way, $J$ and $\vec{J}$ become
functors. Iterating the construction, we can also define the second
order jet bundle $J^2 E$ by (a)~considering the iterated first order
jet bundle~$J(JE)$, (b)~passing to the so-called semiholonomic second
order jet bundle $\bar{J}^2 E$, which is the affine subbundle of~$J(JE)$
over~$JE$ defined by the condition
\begin{equation} \label{eq:SOJB1}
 \bar{J}^2 E~=~\{ \, w \smin J(JE) \, | \;
                     \tau_{JE}(w)~=~J \tau_E(w) \, \}~,
\end{equation}
where $\tau_E$ and $\tau_{JE}$ are the target projections of $JE$ and of~%
$J(JE)$, respectively, while $\, J \tau_E: J(JE) \longrightarrow JE \,$
is the jet prolongation of $\, \tau_E: JE \longrightarrow E$, and (c)~%
decomposing this, as a fiber product of affine bundles over~$JE$, into
a symmetric part and an antisymmetric part: the former is precisely
$J^2 E$ and is an affine bundle over~$JE$, with difference vector bundle
equal to the pull-back to~$JE$ of the vector bundle $\, \pi^* \bigl(
\bvee^{2\,} T^\ast M \bigr) \otimes VE \,$ over~$E$ by the target
projection $\tau_E$, whereas the latter is a vector bundle over~$JE$,
namely the pull-back to~$JE$ of the vector bundle $\, \pi^* \bigl(
\bwedge^{\!2\,} T^\ast M \bigr) \otimes VE \,$ over~$E$ by the target
projection~$\tau_E$:
\begin{equation} \label{eq:SOJB2}
 \begin{array}{c}
  \bar{J}^2 E~\cong~J^2 E \; \times_{JE}^{} \;
                    \tau_E^* \Bigl( \pi^* \bigl( \bwedge^{\!2\,} T^\ast M \bigr)
                                    \otimes VE \Bigr) \\[2mm]
  \vec{J^2} E~\cong~\tau_E^* \Bigl( \pi^* \bigl( \bvee^{2\,} T^\ast M \bigr)
                                    \otimes VE \Bigr)
 \end{array}
\end{equation}

Turning to gauge theories, we begin by recalling that the starting
point for the formulation of a gauge theory is the choice of (a)~a~Lie
group~$G$, with Lie algebra~$\mathfrak{g}$, (b)~a~principal $G$-bundle
$P$ over the space-time manifold~$M$ with projection $\, \rho: P
\longrightarrow M \,$ and carrying a naturally defined right action
of~$G$ that will be written in the form
\begin{equation} \label{eq:ACTGP}
 \begin{array}{ccc}
  P \times G & \longrightarrow &     P     \\[1mm]
    (p\,,g)  &   \longmapsto   & p \cdot g
 \end{array}
\end{equation}
and (c)~a~manifold $Q$ carrying an action of $G$ as in eqn~(\ref{eq:ACTGQ})
above, so that we can form the associated bundle $\, E = P \times_G Q \,$
over~$M$ as well as the connection bundle $\, CP = JP/G$ \linebreak
over~$M$. Sections of~$E$ represent the (multiplet of all) matter
fields present in the theory, whereas sections of $CP$ represent
the gauge potentials (connections). The group $G$ is usually
referred to as the structure group of the model: it contains a
compact ``internal part'' (e.g., a $U(1)$ factor for electrodynamics
or, more generally, an $SU(2) \times U(1)$ factor for the electroweak
theory, an $SU(3)$ factor for chromodynamics, etc.) but possibly also
a non-compact ``space-time part'' which is an appropriate spin group,
in order to accomodate tensor and spinor fields.

The constructions of the associated bundle $\, E = P \times_G Q \,$ and
of the connection bundle $\, CP = JP/G \,$ are standard (both can be
obtained as quotients by a free action of~$G$), and we just recall a
few basic aspects in order to fix notation.

The first arises from the cartesian product $P \times Q$, on which
$G$ acts according to
\[
 g \cdot (p\,,q)~=~(p \cdot g^{-1} , g \cdot q)~,
\]
and we denote the equivalence class (orbit) of a point $\, (p\,,q)
\smin\, P \times Q \,$ by $\, [p\,,q] \smin\, P \times_G Q$.

The second is obtained from the first order jet bundle $JP$ of~$P$,
which carries a natural right $G$-action induced from the given right
$G$-action on~$P$ by taking derivatives (jets) of local sections:
for any point $\, m \smin M$,
\[
 (j_m \sigma) \cdot g~=~j_m(\sigma \cdot g)~,
\]
and we denote the equivalence class (orbit) of a point $\, j_m \sigma
\smin\, JP \,$ by $\, [j_m \sigma] \smin\, CP$. \linebreak Such an
equivalence class can be identified with a horizontal lifting map on
the fiber $P_m$ of~$P$ at~$m$, that is, a $G$-equivariant homomorphism
\[
 \Gamma_m : P_m \times T_m M \longrightarrow TP \, | \, P_m
\]
of vector bundles over~$P_m$ which, when composed with the restriction to~%
$TP \, | \, P_m$ of the tangent map $\, T \rho: TP \longrightarrow TM \,$
to the principal bundle projection $\, \rho: P \longrightarrow M$, gives
the identity. Still another representation, and the most useful one for
practical purposes, is in terms of a connection form on the fiber $P_m$
of~$P$ at~$m$, that is, a $G$-equivariant $1$-form $A_m$ on~$P_m$ with
values in the Lie algebra $\mathfrak{g}$ whose restriction to the
vertical subspace $V_p P$ at each point $p$ of~$P_m$ coincides with
the canonical isomorphism of $V_p P$ with $\mathfrak{g}$ given by
the standard formula for fundamental vector fields,
\[
 \begin{array}{ccc}
  \mathfrak{g} & \longrightarrow &      V_p P      \\[1mm]
        X      &   \longmapsto   & \hat{X}_P^{}(p)
 \end{array} \qquad \mbox{where} \qquad
 \hat{X}_P^{}(p)~=~\frac{d}{dt} \; p \cdot \exp(tX) \, \Big|_{t=0}~,
\]
the relation between the two being that the image of $\Gamma_m$ coincides
with the kernel of $A_m$. Since $JP$ is an affine bundle over~$P$, with
difference vector bundle $\, \vec{J} P \cong \rho^*(T^* M) \otimes VP \,$
where $VP$ is the vertical bundle of~$P$ which in turn is canonically
isomorphic to the trivial vector bundle $P \times \mathfrak{g}$ over~$P$,
and since $G$ acts by fiberwise affine transformations on~$JP$ and by
fiberwise linear transformations on~$\vec{J} P$, it follows that $CP$
is an affine bundle over~$M$, with difference vector bundle
$\, \vec{C} P \cong T^* M \otimes (P \times_G \mathfrak{g})$.

Finally, we shall also need to consider the first order jet bundle of the
connection bundle, which turns out to admit a canonical decomposition into
a symmetric and an antisymmetric part. To derive this, note that when we
lift the given right $G$-action from~$P$ first to~$JP$ as above and then
to~$J(JP)$ by applying the same prescription again, the semiholonomic
second order jet bundle $\bar{J}^2 P$ will be a $G$-invariant subbundle
of~$J(JP)$ and its quotient by~$G$ will be precisely~$J(CP)$:
\begin{equation} \label{eq:JCPJ2P}
 J(CP)~\cong~\bar{J}^2 P / G~.
\end{equation}
But the canonical decomposition of $\bar{J}^2 P$ into symmetric and
antisymmetric part as given in eqn~(\ref{eq:SOJB2}), when specialized
to principal bundles, is manifestly $G$-invariant, so we can divide
by the $G$-action to arrive at the desired canonical decomposition
of $J(CP)$, as a fiber product of affine bundles over~$CP$, into a
symmetric part (an affine bundle) and an antisymmetric part (vector
bundle):
\begin{equation} \label{eq:DECJCP}
 \begin{array}{c}
  J(CP)~\cong~(J^2 P)/G \; \times_{CP}^{} \;
              \pi_{CP}^* \bigl( \bwedge^{\!2\,} T^* M \otimes
                                (P \times_G \mathfrak{g}) \bigr) \\[2mm]
  (\vec{J^2} P)/G~\cong~\pi_{CP}^* \bigl( \bvee^{2\,} T^* M \otimes
                                (P \times_G \mathfrak{g}) \bigr)
 \end{array}
\end{equation}
The projection onto the second summand in this decomposition is called
the \textbf{curvature map} because, at the level of sections, it maps
the $1$-jet of a connection form $A$ to its curvature form $F_A$.
See Theorem~1 of~Ref.~\cite{MM} and Theorems~5.3.4 and~5.3.5 of~%
Ref.~\cite{Sa}.

\section{Gauge group bundles and their actions}

In this section we introduce the basic object needed to describe symmetries
in gauge theories within a purely finite-dimensional setting: the gauge group
bundle and its descendants. All of these are Lie group bundles which admit
certain natural actions on various of the bundles introduced before and/or
their descendants: our aim here will be to define them rigorously.
In order to do so, we must first introduce the notion of a Lie group
bundle and of an action of a Lie group bundle on a fiber bundle (over
the same base manifold): this involves the concept of a locally constant
structure on a fiber bundle.

Let $E$ be a fiber bundle over a base manifold~$M$ with projection
$\, \pi: E \rightarrow M \,$ and typical fiber~$E_0^{}$, and suppose
that both $E_0^{}$ and every fiber $E_m^{}$ of~$E$ are equipped with
some determined geometric structure of the same kind. We say that
such a geometric structure is \textbf{locally constant along $M$}
if there exists a family $(\Phi_\alpha^{})_{\alpha \in A}^{}$ of local
trivializations \linebreak $\Phi_\alpha^{}: \pi^{-1}(U_\alpha^{})
\rightarrow U_\alpha^{} \times E_0^{} \;$ of~$E$ whose domains
$U_\alpha^{}$ cover~$M$ and such that for every point $m$
in~$U_\alpha^{}$ the diffeomorphism $\; (\Phi_\alpha^{})_m^{}:
E_m^{} \rightarrow E_0^{} \;$ is structure preserving: any
(local) trivialization of this kind will be called
\textbf{compatible}.
\begin{dfn}~ \label{def:LGB}
 A \textbf{Lie group bundle} (often abbreviated ``LGB'') over a given base
 mani\-fold\/~$M$ is a fiber bundle\/ $\bar{G}$ over\/~$M$ whose typical
 fiber is a Lie group\/~$G$ and which comes equipped with a strict bundle
 homomorphism over\/~$M$
 \begin{equation} \label{eq:DEFLGB1}
  \begin{array}{ccc}
   \bar{G} \times_M^{} \bar{G} & \longrightarrow & \bar{G} \\[1mm]
              (h,g)            &   \longmapsto   &   h g
  \end{array}
 \end{equation}
 called \textbf{multiplication} or the \textbf{product}, a global section\/
 $1$ of\/~$\bar{G}$ called the \textbf{unit} and a strict bundle homomorphism
 over\/~$M$
 \begin{equation} \label{eq:DEFLGB2}
  \begin{array}{ccc}
   \bar{G} & \longrightarrow & \bar{G} \\[1mm]
      g    &   \longmapsto   &  g^{-1}
  \end{array}
 \end{equation}
 called \textbf{inversion} which, taken together, define a Lie group
 structure in each fiber of\/~$\bar{G}$ that is locally constant
 along\/~$M$.
\end{dfn}
This definition coincides with the one adopted in Ref.~\cite[p.~11]{MK}.
\begin{exe}~ \label{exe:GGB}
 Let~$P$ be a principal bundle over a given manifold~$M$ with structure
 group~$G$ and bundle projection $\; \rho: P \rightarrow M$. Then the
 associated bundle $\, P \times_G G$, where $G$ is supposed to act on
 itself by conjugation, is a Lie group bundle over~$M$, if we define
 multiplication by
 \begin{equation} \label{eq:EXELGB1}
  [p\,,h] [p\,,g]~=~[p\,,h g]~,
 \end{equation}
 the unit by
 \begin{equation} \label{eq:EXELGB2}
  1_{\rho(p)}^{}~=~[p\,,1]~,
 \end{equation}
 and inversion by
 \begin{equation} \label{eq:EXELGB3}
  [p\,,g]^{-1}~=~[p\,,g^{-1}]~.
 \end{equation}
 We call it the \textbf{gauge group bundle} associated with~$P$ because
 the group of its sections is naturally isomorphic to the group of strict
 automorphisms of~$P$, which is usually referred to as the group of gauge
 transformations (or sometimes simply, though somewhat misleadingly, the
 gauge group) associated with~$P$:
 \begin{equation} \label{eq:EXELGB4}
  \Gamma(P \times_G G)~\cong~\mathrm{Aut}_s(P)~.
 \end{equation} 
\end{exe}
\begin{dfn}~ \label{def:ACTLGB}
 An \textbf{action} of a Lie group bundle\/~$\bar{G}$ on a fiber
 bundle\/~$E$, both over the same given base manifold\/~$M$, is a
 strict bundle homomorphism over\/~$M$
 \begin{equation} \label{eq:ACTLGB1}
  \begin{array}{ccc}
   \bar{G} \times_M^{} E & \longrightarrow &     E     \\[1mm]
           (g,e)         &   \longmapsto   & g \cdot e
  \end{array}
 \end{equation}
 which defines an action of each fiber of\/~$\bar{G}$ on the corresponding
 fiber of\/~$E$ that is locally constant along\/~$M$.
\end{dfn}
\begin{exe}~ \label{exe:ACTGGB1}
 Let~$P$ be a principal bundle over a given manifold~$M$ with structure
 group~$G$ and bundle projection $\; \rho: P \rightarrow M \;$ and $Q$
 be a manifold carrying an action of $G$ as in eqn~(\ref{eq:ACTGQ})
 above. Then the gauge group bundle $P \times_G G$ acts naturally on
 the associated bundle $P \times_G Q$, according to
 \begin{equation} \label{eq:ACTGGB1}
  [p\,,g] \cdot [p\,,q]~=~[p\,,g \cdot q]~.
 \end{equation}
 A particular case occurs if we take $Q$ to be $G$ itself, but this time
 letting $G$ (the structure group) act on~$G$ (the typical fiber) by left
 translation: using the fact that the resulting associated bundle is
 canonically isomorphic to~$P$ itself,%
 \footnote{Explicitly, this isomorphism is given by mapping $[p,g]$ to~$p$.}
 we see that the gauge group bundle $P \times_G G$ acts naturally on~$P$
 itself, according to
 \begin{equation} \label{eq:ACTGGB2}
  [p\,,g] \cdot (p \cdot g_0)~=~p \cdot (g g_0)~.
 \end{equation}
 Note that this (left) action of~$P \times_G G$ on~$P$ commutes with the
 (right) action of~$G$ on~$P$ specified in eqn~(\ref{eq:ACTGP}) above~--
 a fact which can be viewed as a natural generalization, from Lie groups to
 principal bundles and bundles of Lie groups, of the well known statement
 that left translations commute with right translations (and to which it
 reduces when $M$ is a single point).
\end{exe}

Bundles of Lie groups and their actions also behave naturally under taking
derivatives. For example, we have the following
\begin{prp} \label{prp:ACIN1}~
 An action of a Lie group bundle\/ $\bar{G}$ on a fiber bundle\/~$E$, both
 over the same base manifold\/~$M$, induces in a natural way actions of\/
 $\bar{G}$ on the vertical bundle\/ $V\!E$ of\/~$E$ and on the linearized
 jet bundle\/ $\vec{J} E$ of\/~$E$.
\end{prp}
Explicitly, the first of these induced actions is defined by
\begin{equation}
 g \cdot \frac{d}{dt} \, e(t) \Big|_{t=0}~
 =~\frac{d}{dt} \,  \bigl( g \cdot e(t) \bigr) \Big|_{t=0} \qquad
 \mbox{for $\, m \smin M$, $g \smin \bar{G}_m^{}$,
       $e$ any smooth curve in $E_m^{}$}~,
\end{equation}
whereas the second is obtainded by taking its tensor product with the trivial
action of~$\bar{G}$ on the cotangent bundle $T^* M$ of~$M$, using the canonical
isomorphism $\, \vec{J} E \cong \pi^*(T^* M) \otimes V\!E \,$.
In a slightly different direction, we note the following
\begin{thm} \label{thm:LGBJ1}~
 Let\/ $\bar{G}$ be a Lie group bundle over\/~$M$. Then for any
 $\, r \geqslant 1$, its\/ $r^{\mathrm{th}}$ order jet bundle\/
 $J^r \bar{G}$ is also a Lie group bundle over\/~$M$.%
 \footnote{One should note that an analogous statement for principal
 bundles would be false: the fact that $P$ is a principal bundle does
 not imply that $J^r P$ is a principal bundle.}
 (When $\, r=1$, we often omit the superfix\/~$r$.)
\end{thm}
\begin{thm} \label{thm:LGBJ2}~
 An action of a Lie group bundle\/ $\bar{G}$ on a fiber bundle\/~$E$, both
 over the same base manifold\/~$M$, induces in a natural way an action of\/
 $J^r \bar{G}$ on\/~$J^r E$, for any $\, r \geqslant 1$.
 (When $\, r=1$, we often omit the superfix\/~$r$.)
\end{thm}
\proof
 Both theorems are direct consequences of the fact that taking $r^{\mathrm{th}}$
 order jets is a functor, together with the fact that for any two fiber bundles
 $E$ and~$F$ over the same base manifold~$M$, there is a canonical isomorphism
 \[
  J^r(E \times_M F)~\cong~J^r E \times_M J^r F
 \]
 induced by the identification of (local) sections of $E \times_M F$ with
 pairs of (local) sections of~$E$ and of~$F$. More specifically, we define
 the product, the unit and the inversion in $J^r \bar{G}$ by extending the
 product, the unit and the inversion in $\bar{G}$, respectively, pointwise
 to local sections and then taking $r^{\mathrm{th}}$ order jets, and similarly,
 we define the action of~$J^r \bar{G}$ on $J^r E$ by extending the action
 of~$\bar{G}$ on~$E$ pointwise to local sections and then taking
 $r^{\mathrm{th}}$ order jets.
\qed
\begin{exe}~ \label{exe:DGGB}
 Let~$P$ be a principal bundle over a given manifold~$M$ with structure
 group~$G$ and bundle projection $\; \rho: P \rightarrow M$. Then for any
 $\, r \geqslant 1$, the $r^{\mathrm{th}}$ order jet bundle $J^r(P \times_G G)$
 of the gauge group bundle $\, P \times_G G \,$ is a Lie group bundle over~$M$
 which we shall call the ($r^{\mathrm{th}}$ order) \textbf{derived gauge group
 bundle} associated with~$P$. (When $\, r=1$, we often omit the prefix
 ``first order''.)
\end{exe}

With these tools at our disposal, we proceed to define various actions of
the gauge group bundle and the (first and second order) derived gauge group
bundles that play an important role in the analysis of symmetries in gauge
theories. Starting with the action
\begin{equation} \label{eq:ACTPGPQ}
 (P \times_G G) \times (P \times_G Q)~\longrightarrow~P \times_G Q
\end{equation}
of the gauge group bundle $P \times_G G$ on the matter field bundle
$P \times_G Q$ already mentioned in Example~\ref{exe:ACTGGB1} (cf.\
eqn~(\ref{eq:ACTGGB1})), consider
first the induced action
\begin{equation} \label{eq:ACTPGVPQ}
 (P \times_G G) \times V(P \times_G Q)~\longrightarrow~V(P \times_G Q)
\end{equation}
of $P \times_G G$ on the vertical bundle $V(P \times_G Q)$ of~$P \times_G Q$
obtained by taking derivatives along the fibers (this corresponds to the
transition from the action~(\ref{eq:ACTGQ}) to the action~(\ref{eq:ACTGTQ}),
using that, for each point $m$ in~$M$, the fiber of~$V(P \times_G Q)$ over~$m$
is precisely the tangent bundle of the fiber of~$P \times_G Q$ over~$m$), and
extend it trivially to an action
\begin{equation} \label{eq:ACTPGLJPQ}
 (P \times_G G) \times \vec{J}(P \times_G Q)~\longrightarrow~
 \vec{J}(P \times_G Q)
\end{equation}
of $P \times_G G$ on the linearized first order jet bundle $\vec{J}%
(P \times_G Q)$ of~$P \times_G Q$, using the canonical isomorphism
\[
 \vec{J}(P \times_G Q)~\cong~\pi^*(T^* M) \otimes V(P \times_G Q)
\]
and taking the tensor product with appropriate identities on $\pi^*(T^* M)$.
Similarly, the action~(\ref{eq:ACTPGPQ}) induces an action
\begin{equation} \label{eq:ACTJPGJPQ}
 J(P \times_G G) \times J(P \times_G Q)~\longrightarrow~J(P \times_G Q)
\end{equation}
of $J(P \times_G G)$ on the first order jet bundle $J(P \times_G Q)$
of~$P \times_G Q$, obtained by applying Theorem~\ref{thm:LGBJ2}
(with $\, r=1$). On the other hand, starting with the action
\[
 (P \times_G G) \times P~\longrightarrow~P
\]
of the gauge group bundle $P \times_G G$ on the principal bundle $P$ itself
already mentioned in Example~\ref{exe:ACTGGB1} (cf.\ eqn~(\ref{eq:ACTGGB1})),
which commutes with the (right) action of~$G$ on~$P$, Theorem~\ref{thm:LGBJ2}
with $\, r=1$) provides an action
\[
 J(P \times_G G) \times JP~\longrightarrow~JP
\]
of~$J(P \times_G G)$ on the first order jet bundle $JP$ of~$P$ which
commutes with the induced (right) action of~$G$ on~$JP$ and therefore
factors through the quotient to yield an action
\begin{equation} \label{eq:ACTJPGCP}
 J(P \times_G G) \times CP~\longrightarrow~CP
\end{equation}
of~$J(P \times_G G)$ on the connection bundle~$CP$ of~$P$. Finally,
applying Theorem~\ref{thm:LGBJ2} (with $\, r=1$) once more and using
the fact that $J^2(P \times_G G)$ can be realized as a Lie group
subbundle of the iterated Lie group bundle $J(J(P \times_G G))$,
we arrive at an action
\begin{equation} \label{eq:ACTJ2PGJCP}
 J^2(P \times_G G) \times J(CP)~\longrightarrow~J(CP)
\end{equation}
of~$J^2(P \times_G G)$ on the first order jet bundle $J(CP)$ of~$CP$.

\section{Local expressions}

Our next goal will be to derive local expressions for the actions
(\ref{eq:ACTPGPQ}-\ref{eq:ACTJ2PGJCP}) to show that they are global
versions of well known and intuitively obvious constructions used by
physicists.

In order to build this bridge, the first obstacle to be overcome
is the fact that physicists usually write these actions in terms
of fields, that is, of (local) sections of the bundles involved,
rather than the bundles themselves. In mathematical terms, this
corresponds to thinking in terms of sheaves, rather than bundles.
Now a Lie group bundle corresponds to a ``locally constant'' sheaf
of Lie groups and an action of a Lie group bundle on a fiber bundle
corresponds to a ``locally constant'' action of a ``locally constant''
sheaf of Lie groups on a ``locally constant'' sheaf of manifolds.
Explicitly, these sheaves are obtained by associating with each open
subset $U$ of the base manifold~$M$ the group $\Gamma(U,\bar{G})$ of
sections $\mathsl{g}$ of~$\bar{G}$ over $U$ and the space $\Gamma(U,E)$
of sections $\varphi$ of~$E$ over~$U$, and with each pair of open sub%
sets of~$M$, one of which is contained in the other, the appropriate
restriction maps. The requirement of local triviality then means that
each point of~$M$ admits an open neighborhood such that for every
open subset $U$ of~$M$ contained in it, we have isomorphisms
$\, \Gamma(U,\bar{G}) \cong C^\infty(U,G) \,$ and $\, \Gamma(U,E)
\cong C^\infty(U,Q) \,$, and the requirement of local constancy
means that these isomorphisms can be chosen such that the product,
the unit and the inversion in~$\Gamma(U,\bar{G})$ correspond to the
pointwise defined product, unit and inversion in~$C^\infty(U,G)$ and
the action of $\Gamma(U,\bar{G})$ on~$\Gamma(U,E)$ corresponds to
the pointwise defined action of $C^\infty(U,G)$ on~$C^\infty(U,Q)$:
\begin{eqnarray*}
 &(\mathsl{g}_1 \mathsl{g}_2)(x)~=~\mathsl{g}_1(x) \, \mathsl{g}_2(x)~~~,~~~
 \mathsl{g}^{-1}(x)~=~(\mathsl{g}(x))^{-1}
 \qquad \mbox{for $\,x \smin U$}& \\[2mm]
 &(\mathsl{g} \cdot \varphi)(x)~=~\mathsl{g}(x) \cdot \varphi(x)
 \qquad \mbox{for $\,x \smin U$}&
\end{eqnarray*}
This interpretation is particularly useful for considering induced actions
of jet bundles of~$\bar{G}$ on jet bundles of~$E$, because it allows to
state their definition in the simplest possible way: the action of~%
$J \bar{G}$ on~$JE$, say, induced by an action of~$\bar{G}$ on~$E$,
when translated from an action of fibers on fibers to an action of
local sections on local sections, is simply given by taking the
derivative, according to the standard rules.

Let us apply this strategy to the actions of the gauge group bundle and
its descendants introduced at the end of the previous section. To this
end, we assume throughout the rest of this section that $U$ is an
arbitrary but fixed coordinate domain in~$M$ over which $P$ is
trivial and that we have chosen a section $\sigma$ of~$P$ over~$U$,
together with a system of coordinates $x^\mu$ on~$U$: together, these
will induce, for each of the bundles appearing in eqns~(\ref{eq:ACTPGPQ}-%
\ref{eq:ACTJ2PGJCP}), a trivialization over~$U$ which in turn provides
an isomorphism between its space of sections over~$U$ and an appropriate
function space.%
\footnote{Among these function spaces we will find spaces of the form
$\mathrm{Hom}(E,F)$, where $E$ and~$F$ are vector bundles over (possibly
different) manifolds $M$ and~$N$, respectively, consisting of all vector
bundle homomorphisms from $E$ to~$F$, that is, of all smooth maps from
the manifold~$E$ to the manifold~$F$ which are fiber preserving and
fiberwise linear. As an example, note that for $f$ in~$C^\infty(M,N)$,
its tangent map $Tf$, which we shall also denote by $\partial f$, is
in~$\mathrm{Hom}(TM,TN)$.}

Under the identifications $\, \Gamma(U,P \times_G G) \cong C^\infty(U,G)$,
$\, \Gamma(U,P \times_G Q) \cong C^\infty(U,Q)$, $\, \Gamma(U,V(P \times_G Q))
\cong C^\infty(U,TQ) \,$ and $\, \Gamma(U,\vec{J}(P \times_G Q)) \cong
\mathrm{Hom}(TU,TQ)$, for example, the actions (\ref{eq:ACTPGPQ}),
(\ref{eq:ACTPGVPQ}) and~(\ref{eq:ACTPGLJPQ}) correspond, respectively,
to an action
\begin{equation} \label{eq:ACTPGPQ2}
 \begin{array}{ccc}
  C^\infty(U,G) \times C^\infty(U,Q) & \longrightarrow &
  C^\infty(U,Q) \\[1mm]
         (\mathsl{g},\varphi)        &   \longmapsto   &
  \mathsl{g} \cdot \varphi
 \end{array}
\end{equation}
defined pointwise, i.e., by
\begin{equation} \label{eq:ACTPGPQ3}
 (\mathsl{g} \cdot \varphi)(x)~=~\mathsl{g}(x) \cdot \varphi(x)
 \qquad \mbox{for $\,x \smin U$}~,
\end{equation}
as above, where the symbol $\cdot$ on the rhs stands for the original action
(\ref{eq:ACTGQ}) of~$G$ on~$Q$, to an action
\begin{equation} \label{eq:ACTPGVPQ2}
 \begin{array}{ccc}
  C^\infty(U,G) \times C^\infty(U,TQ) & \longrightarrow &
  C^\infty(U,TQ) \\[1mm]
      (\mathsl{g},\delta\varphi)      &   \longmapsto   &
  \mathsl{g} \cdot \delta\varphi
 \end{array}
\end{equation}
defined pointwise, i.e., by
\begin{equation} \label{eq:ACTPGVPQ3}
 (\mathsl{g} \cdot \delta\varphi)(x)~=~\mathsl{g}(x) \cdot \delta\varphi(x)
 \qquad \mbox{for $\,x \smin U$}~,
\end{equation}
and to an action
\begin{equation} \label{eq:ACTPGLJPQ2}
 \begin{array}{ccc}
  C^\infty(U,G) \times \mathrm{Hom}(TU,TQ) & \longrightarrow &
  \mathrm{Hom}(TU,TQ) \\[1mm]
           (\mathsl{g},D\varphi)           &   \longmapsto   &
  \mathsl{g} \cdot D\varphi
 \end{array}
\end{equation}
defined pointwise, i.e., by
\begin{equation} \label{eq:ACTPGLJPQ3}
 (\mathsl{g} \cdot D\varphi)(x,u)~
 =~\mathsl{g}(x) \cdot D\varphi(x,u)
 \qquad \mbox{for $\,x \smin U$, $u \smin T_x M$}~,
\end{equation}
where the symbol $\cdot$ on the rhs now stands for the induced action
(\ref{eq:ACTGTQ}) of~$G$ on~$TQ$ and we have used the symbols $\delta
\varphi$ and $D\varphi$ to indicate that the transformation laws of
these objects correspond to those of variations of~sections\,%
\footnote{Variations $\delta\varphi$ of a section $\varphi$ of a fiber
bundle $E$ over~$M$ are formal first order derivatives of one-parameter
families $(\varphi_s)_{-\epsilon<s<\epsilon}$ of sections of~$E$ around
$\varphi$ with respect to the parameter ($\varphi_0=\varphi$,
$d\varphi_s/ds\,|_{s=0}=\delta\varphi$) and are therefore sections
of the pull-back $\varphi^*(VE)$ of the vertical bundle $VE$ of~$E$
to~$M$ via~$\varphi$. If, formally, one considers the space $\Gamma(E)$
of sections of~$E$ as a manifold, they form its tangent space at~$\varphi$:
$T_\varphi(\Gamma(E)) = \Gamma(\varphi^*(VE))$.}
and of covariant derivatives of~sections\,%
\footnote{Given an arbitrary connection in a fiber bundle $E$ over~$M$,
which can be viewed as the choice of a horizontal bundle, or equivalently,
a vertical projection, or equivalently, a horizontal projection, one has
a notion of covariant derivative: the covariant derivative of a section
is obtained from its ordinary derivative (tangent map) by composition
with the vertical projection. Thus, for any section $\varphi$ of~$E$
and any point $\, m \ssmin M$, its ordinary derivative at~$m$ can be
viewed as belonging to the affine space $J_{\varphi(m)} E$ (jet space)
but its covariant derivative at~$m$ belongs to the vector space
$\vec{J}_{\varphi(m)} E$ (linearized jet space).}.
Similarly, under the identifications $\, \Gamma(U,J(P \times_G G))
\cong \mathrm{Hom}(TU,TG) \,$ and \linebreak $\Gamma(U,J(P \times_G Q))
\cong \mathrm{Hom}(TU,TQ)$, the action~(\ref{eq:ACTJPGJPQ}) corresponds
to an action
\begin{equation} \label{eq:ACTJPGJPQ2}
 \begin{array}{ccc}
  \mathrm{Hom}(TU,TG) \times \mathrm{Hom}(TU,TQ) & \longrightarrow &
  \mathrm{Hom}(TU,TQ) \\[1mm]
       (\partial\mathsl{g},\partial\varphi)      &   \longmapsto   &
  \partial\mathsl{g} \cdot \partial\varphi
 \end{array}
\end{equation}
defined pointwise, i.e., by
\begin{equation} \label{eq:ACTJPGJPQ3}
 (\partial\mathsl{g} \cdot \partial\varphi)(x,u)~
 =~\partial\mathsl{g}(x,u) \cdot \partial\varphi(x,u)
 \qquad \mbox{for $\,x \smin U$, $u \smin T_x M$}~,
\end{equation}
where the symbol $\cdot$ on the rhs now stands for the induced
action (\ref{eq:ACTTGTQ}) of~$TG$ on~$TQ$ and we have used the
symbol $\partial\varphi$ to indicate that the transformation law
of this object corresponds to that of ordinary derivatives of
sections; we can also imagine the symbol $\partial\varphi$ to
represent the collection of all partial derivatives $\partial_\mu
\varphi$ of~$\varphi$ and, similarly, the symbol $\partial\mathsl{g}$
to represent the collection of all partial derivatives $\partial_\mu
\mathsl{g}$ of $\mathsl{g}$.

In order to deal with the remaining two actions, as well as to rewrite
the last of the previous ones in a different form, we use additional
identifications provided by the trivialization~(\ref{eq:TRITG}) of
the tangent bundle $TG$ of~$G$ and by the interpretation of sections
of~$CP$ as connection forms to represent the pertinent objects in a
more manageable form. Writing $\, \mathscr{T}_1^0(U,\mathfrak{g})
= \Omega^1(U,\mathfrak{g})$ for the space of rank~$1$ tensor fields,
or $1$-forms, $\mathscr{T}_2^0(U,\mathfrak{g})$ for the space of
rank~$2$ tensor fields and $\mathscr{T}_{2,s}^0(U,\mathfrak{g})$
for the space of symmetric rank~$2$ tensor fields on~$U$ with
values in~$\mathfrak{g}$, the pertinent isomorphisms for the
derived gauge group bundles are
\[
 \Gamma(U,J(P \times_G G))~\cong~\mathrm{Hom}(TU,TG)~
 \cong~\mathrm{Hom}(TU,G \times \mathfrak{g})~
 \cong~C^\infty(U,G) \times \mathscr{T}_1^0(U,\mathfrak{g})~,
\]
and
\[
 \Gamma(U,J^2(P \times_G G))~
 \cong~C^\infty(U,G) \times \mathscr{T}_1^0(U,\mathfrak{g}) \times
                            \mathscr{T}_{2,s}^0(U,\mathfrak{g})~,
\]
while those for the connection bundle and its first order jet bundle
are
\[
 \Gamma(U,CP) \cong \mathscr{T}_1^0(U,\mathfrak{g})~,
\]
and
\[
 \Gamma(U,J(CP)) \cong \mathscr{T}_1^0(U,\mathfrak{g}) \times
 \mathscr{T}_2^0(U,\mathfrak{g})~,
\]
respectively. Explicitly, in terms of a $G$-valued function $\mathsl{g}$
on~$U$ representing a section of $P \times_G G$ over~$U$, the first two
isomorphisms are given by the prescription of taking its $1$-jet,
symbolically represented by the pair $(\mathsl{g},\partial\mathsl{g})$,
to the pair $(\mathsl{g},\partial\mathsl{g} \, \mathsl{g}^{-1})$ and its
$2$-jet, symbolically represented by the triple $(\mathsl{g},\partial
\mathsl{g},\partial^{\,2} \mathsl{g})$, to the triple $(\mathsl{g},
\partial\mathsl{g} \, \mathsl{g}^{-1},\partial (\partial\mathsl{g} \,
\mathsl{g}^{-1}))$, where we can imagine the expression $\partial^{\,2}
\mathsl{g}$ to represent the collection of all second order partial
derivatives $\partial_\mu \partial_\nu \varphi$ of~$\varphi$ while
the expression $\partial (\partial\mathsl{g} \, \mathsl{g}^{-1})$ is
supposed to represent the collection of symmetrized partial derivatives
$\partial_{(\mu} (\partial_{\nu)} \mathsl{g} \, \mathsl{g}^{-1})$,
with expansions into components as follows:
\[
 \begin{array}{c}
  \partial\mathsl{g} \mathsl{g}^{-1}~
  =~(\partial_\mu \mathsl{g} \mathsl{g}^{-1})~dx^\mu \\[2mm]
  \partial (\partial\mathsl{g} \, \mathsl{g}^{-1})~
  =~{\textstyle \frac{1}{2}}
    \bigl( \partial_\mu (\partial_\nu \mathsl{g} \mathsl{g}^{-1}) +
           \partial_\nu (\partial_\mu \mathsl{g} \mathsl{g}^{-1}) \bigr)~
    dx^\mu \otimes dx^\nu
 \end{array}~.
\]
Similarly, the last two isomorphisms amount to representing a section
of~$CP$ over~$U$ by a $\mathfrak{g}$-valued connection $1$-form $A$
on~$U$ and its $1$-jet by the pair $(A,\partial A)$, where we can
imagine the expression $\partial A$ to represent the collection of
all partial derivatives $\partial_\mu A_\nu$, with expansions into
components as follows:
\[
 \begin{array}{c}
  A~=~A_\mu~dx^\mu \\[2mm]
  \partial A~=~\partial_\mu A_\nu~dx^\mu \otimes dx^\nu
 \end{array}~.
\]
With these identifications, the action~(\ref{eq:ACTJPGJPQ2}) can be
rewritten as an action
\begin{equation} \label{eq:ACTJPGJPQ4}
 \begin{array}{ccc}
  (C^\infty(U,G) \times \mathscr{T}_1^0(U,\mathfrak{g}))
 \times \mathrm{Hom}(TU,TQ)
  & \longrightarrow & \mathrm{Hom}(TU,TQ) \\[1mm]
  ((\mathsl{g},\partial\mathsl{g} \mathsl{g}^{-1}),\partial\varphi)
  &   \longmapsto   &
  (\mathsl{g},\partial\mathsl{g} \mathsl{g}^{-1}) \cdot \partial\varphi
 \end{array}
\end{equation}
defined pointwise, i.e., by
\begin{equation} \label{eq:ACTJPGJPQ5}
 \begin{array}{c}
  ((\mathsl{g},\partial\mathsl{g} \mathsl{g}^{-1})
  \cdot \partial\varphi)(x,u)~
  =~\mathsl{g}(x) \cdot (\partial\varphi(x,u)) \, + \,
    \bigl( (\partial\mathsl{g} \mathsl{g}^{-1})(x,u) \bigr)_Q^{}%
    \bigl( \mathsl{g}(x) \cdot \varphi(x) \bigr) \\[1mm]
  \mbox{for $\,x \smin U$, $u \smin T_x M$}
 \end{array}~,
\end{equation}
where the second symbol $\cdot$ on the rhs stands for the original action
(\ref{eq:ACTGQ}) of~$G$ on~$Q$, the first symbol $\cdot$ on the rhs for
the induced action (\ref{eq:ACTGTQ}) of~$G$ on~$TQ$ and, for any $\, X
\smin \mathfrak{g}$, $X_Q$ denotes the fundamental vector field on~$Q$
associated with this generator: this is the precise meaning of the
intuitive but formal ``Leibniz rule for group actions'':
\[
 \partial(\mathsl{g} \cdot \varphi)~
 =~\mathsl{g} \cdot \partial\varphi + \partial\mathsl{g} \cdot \varphi~.
\]
Similarly, the action~(\ref{eq:ACTJPGCP}) corresponds to an action
\begin{equation} \label{eq:ACTJPGCP2}
  \bigl( C^\infty(U,G) \times \mathscr{T}_1^0(U,\mathfrak{g}) \bigr)
  \times \mathscr{T}_1^0(U,\mathfrak{g})~
  \longrightarrow~\mathscr{T}_1^0(U,\mathfrak{g})
\end{equation}
taking $\, \bigl( \bigl( \mathsl{g} , \partial\mathsl{g} \mathsl{g}^{-1}
\bigr) , A \bigr) \,$ to $\bigl( \mathsl{g} , \partial\mathsl{g}
\mathsl{g}^{-1} \bigr) \cdot A$, which is given by the well known
transformation law
\begin{equation} \label{eq:ACTJPGCP3}
 \bigl( \bigl( \mathsl{g} , \partial\mathsl{g} \mathsl{g}^{-1} \bigr)
        \cdot A \bigr)_\mu~
 =~\mathsl{g} A_\mu \, \mathsl{g}^{-1} \, - \,
   \partial_\mu \mathsl{g} \mathsl{g}^{-1}~,
\end{equation}
and finally the action (\ref{eq:ACTJ2PGJCP}) corresponds to an action
\begin{equation} \label{eq:ACTJ2PGJCP2}
  \bigl( C^\infty(U,G) \times \mathscr{T}_1^0(U,\mathfrak{g}) \times
  \mathscr{T}_{2,s}^0(U,\mathfrak{g}) \bigr) \times
  \bigl( \mathscr{T}_1^0(U,\mathfrak{g}) \times
         \mathscr{T}_2^0(U,\mathfrak{g}) \bigr)~
  \longrightarrow~
  \mathscr{T}_1^0(U,\mathfrak{g}) \times
  \mathscr{T}_2^0(U,\mathfrak{g})
\end{equation}
taking $\, \bigl( \bigl( \mathsl{g} , \partial\mathsl{g} \mathsl{g}^{-1} ,
\partial (\partial\mathsl{g} \, \mathsl{g}^{-1}) \bigr) , \bigl( A ,
\partial A \bigr) \bigr) \,$ to $\bigl( \mathsl{g} , \partial\mathsl{g}
\mathsl{g}^{-1} , \partial (\partial\mathsl{g} \, \mathsl{g}^{-1}) \bigr)
\cdot \bigl( A , \partial A \bigr)$, which is given by differentiating
eqn~(\ref{eq:ACTJPGCP3}) and rearranging terms:
\begin{equation} \label{eq:ACTJ2PGJCP3}
 \begin{array}{l}
  \bigl( \bigl( \mathsl{g} , \partial\mathsl{g} \mathsl{g}^{-1} ,
                \partial (\partial\mathsl{g} \, \mathsl{g}^{-1}) \bigr)
         \cdot \bigl( A , \partial A \bigr) \bigr)_{\mu\nu} \\[2mm]
  \mbox{}~~=~\mathsl{g} \, \partial_\mu A_\nu \, \mathsl{g}^{-1} \, + \,
             [ \partial_\mu \mathsl{g} \mathsl{g}^{-1} \,,
               \mathsl{g} A_\nu \mathsl{g}^{-1} ] \, - \,
             \frac{1}{2} \; [ \partial_\mu \mathsl{g} \mathsl{g}^{-1} \,,
                              \partial_\nu \mathsl{g} \mathsl{g}^{-1} ] \\[2mm]
  \mbox{} \qquad - \,
          \frac{1}{2} \; \partial_\mu
          (\partial_\nu \mathsl{g} \mathsl{g}^{-1}) \, - \,
          \frac{1}{2} \; \partial_\nu
          (\partial_\mu \mathsl{g} \mathsl{g}^{-1})~.
 \end{array}
\end{equation}
For later use, we decompose this transformation law into its symmetric part
\begin{equation} \label{eq:ACTJ2PGJCP4}
 \begin{array}{l}
  \bigl( \bigl( \mathsl{g} , \partial\mathsl{g} \mathsl{g}^{-1} ,
                \partial (\partial\mathsl{g} \, \mathsl{g}^{-1}) \bigr)
         \cdot \bigl( A , \partial A \bigr) \bigr)_{\mu\nu} \, + \,
  \bigl( \bigl( \mathsl{g} , \partial\mathsl{g} \mathsl{g}^{-1} ,
                \partial (\partial\mathsl{g} \, \mathsl{g}^{-1}) \bigr)
         \cdot \bigl( A , \partial A \bigr) \bigr)_{\nu\mu} \\[2mm]
  \mbox{}~~=~\mathsl{g} \bigl( \partial_\mu A_\nu + \partial_\nu A_\mu \bigr) \,
             \mathsl{g}^{-1} \, + \,
             [ \partial_\mu \mathsl{g} \mathsl{g}^{-1} \,,
               \mathsl{g} A_\nu \mathsl{g}^{-1} ] \, + \,
             [ \partial_\nu \mathsl{g} \mathsl{g}^{-1} \,,
               \mathsl{g} A_\mu \mathsl{g}^{-1} ] \\[2mm]
  \mbox{} \qquad - \,
          \frac{1}{2} \; \partial_\mu
          (\partial_\nu \mathsl{g} \mathsl{g}^{-1}) \, - \,
          \frac{1}{2} \; \partial_\nu
          (\partial_\mu \mathsl{g} \mathsl{g}^{-1})~,
 \end{array}
\end{equation}
and its antisymmetric part
\begin{equation} \label{eq:ACTJ2PGJCP5}
 \begin{array}{l}
  \bigl( \bigl( \mathsl{g} , \partial\mathsl{g} \mathsl{g}^{-1} ,
                \partial (\partial\mathsl{g} \, \mathsl{g}^{-1}) \bigr)
         \cdot \bigl( A , \partial A \bigr) \bigr)_{\mu\nu} \, - \,
  \bigl( \bigl( \mathsl{g} , \partial\mathsl{g} \mathsl{g}^{-1} ,
                \partial (\partial\mathsl{g} \, \mathsl{g}^{-1}) \bigr)
         \cdot \bigl( A , \partial A \bigr) \bigr)_{\nu\mu} \\[2mm]
  \mbox{}~~=~\mathsl{g} \bigl( \partial_\mu A_\nu - \partial_\nu A_\mu \bigr) \,
             \mathsl{g}^{-1} \, + \,
             [ \partial_\mu \mathsl{g} \mathsl{g}^{-1} \,,
               \mathsl{g} A_\nu \mathsl{g}^{-1} ] \, - \,
             [ \partial_\nu \mathsl{g} \mathsl{g}^{-1} \,,
               \mathsl{g} A_\mu \mathsl{g}^{-1} ] \\[2mm]
  \mbox{} \qquad - \,
          [ \partial_\mu \mathsl{g} \mathsl{g}^{-1} \,,
            \partial_\nu \mathsl{g} \mathsl{g}^{-1} ]~,
 \end{array}
\end{equation}
the latter being equivalent to the simple transformation law
\begin{equation} \label{eq:ACTJ2PGJCP6}
 \bigl( \mathsl{g} \cdot F \bigr)_{\mu\nu}~
 =~\mathsl{g} F_{\mu\nu} \mathsl{g}^{-1}
\end{equation}
for the curvature form $F$ of the connection form~$A$, with components
\[
 F_{\mu\nu} = \partial_\mu A_\nu - \partial_\nu A_\mu + [A_\mu,A_\nu]~.
\]
As is well known, the affine transformation law~(\ref{eq:ACTJPGCP3}) implies
that one can, at any given point of~$M$, gauge the potential $A$ to zero by
an appropriate choice of gauge transformation, namely by assuming that
at the given point $x$, the value of~$\mathsl{g}$ is $1$ and that of its
first order partial derivatives is given by $\, \partial_\mu \mathsl{g}(x)
= A_\mu(x)$. In the language adopted here, this translates into
\begin{prp}~ \label{prp:ACTJPGCP}
 The action of\/~$J(P \times_G G)$ on\/~$CP$ is fiber transitive.
\end{prp}
Similarly, the affine transformation laws~(\ref{eq:ACTJ2PGJCP3}-%
\ref{eq:ACTJ2PGJCP5}) imply that one can, at any given point of~$M$,
gauge the potential $A$ to zero, the symmetric part of its derivative
to zero and the antisymmetric part of its derivative to be equal to
its curvature by an appropriate choice of gauge transformation,
namely by assuming that at the given point $x$, the value of~%
$\mathsl{g}$ is $1$, that of its first order partial derivatives
is given by $\, \partial_\mu \mathsl{g}(x) = A_\mu(x) \,$ and
that of its second order partial derivatives is given by
$\; \bigl( \partial_\mu (\partial_\nu \mathsl{g} \, \mathsl{g}^{-1})
+ \partial_\nu (\partial_\mu \mathsl{g} \, \mathsl{g}^{-1}) \bigr)(x)
=  \bigl( \partial_\mu A_\nu + \partial_\nu A_\mu \bigr)(x) \,$.
In the language adopted here, this translates into
\begin{prp}~ \label{prp:ACTJ2PGJCP}
 The action of\/~$J^2(P \times_G G)$ on\/~$J(CP)$ preserves the decomposition~%
 (\ref{eq:DECJCP}) into symmetric and antisymmetric part, is fiber transitive
 on the symmetric part and under the curvature map (which is the projection
 to the antisymmetric part) is taken to the natural action of\/~$P \times_G G$
 on\/~$\, \bwedge^{\!2\,} T^* M \otimes (P \times_G \mathfrak{g}) \,$ induced
 by the adjoint representation of\/~$G$ on\/~$\mathfrak{g}$.
\end{prp}

\section{Global and local invariance}

In the usual geometric formulation of a gauge theory with structure group~$G$
and under\-lying principal $G$-bundle $P$ over space-time~$M$, gauge potentials
are represented by connection forms $A$ on~$P$, which can be reinterpreted as
sections of the bundle $CP$ of principal connections on~$P$, and the matter
fields (all assembled into one big multiplet) by sections of a fiber bundle
$\, P \times_G Q \,$ associated to~$P$. The configuration bundle of the whole
theory is thus the fiber product $\, E = CP \times_M (P \times_G Q)$. The group
$\mathrm{Aut}(P)$ of automorphisms of~$P$ and the subgroup $\mathrm{Gau}(P)$
of gauge transformations, or in mathematical terms, of strict automorphisms
of~$P$, act naturally on~$\Gamma(CP)$ and on~$\Gamma(P \times_G Q)$ and,
therefore, also on~$\Gamma(E)$, and it is to this group and its actions
that one usually refers to when speaking about gauge invariance.
Unfortunately, the group $\mathrm{Gau}(P)$ and all the spaces on
which it acts are infinite-dimensional, which makes this kind of
symmetry very hard to handle. As~observed in the introduction,
this happens already in the case of mechanics. And~precisely as in that
case, there is a way out: we can convert the principle of gauge invariance
into a completely finite-dimensional setting by making use of the natural
actions of Lie group bundles introduced in the preceding two sections.
In fact, in a completely geometric formulation of field theory, even
the correct implementation of global symmetries, in the matter field
sector, already requires the use of actions of Lie group bundles over
space-time, rather than just ordinary Lie groups!

At first sight, the reader will probably find the last statement rather
surprising (we certainly did when we first stumbled over it), partly because
it seems to go against wide\-spread belief (but as the reader will be able
to convince himself as we go along, this is not really the case), partly
also because it is not immediately visible in the mechanical analogue
discussed in the introduction. However, its origins are really quite easy
to understand. Consider, for example, a function $V$ on the total space
$P \times_G Q$, which may represent a potential term for some matter field
lagrangian: then since it is not the Lie group $G$  itself but rather the
Lie group bundle $P \times_G G$ that acts naturally on this total space,%
\footnote{Recall that in the definition of $P \times_G Q$, we divide
$P \times Q$ by the joint action of $G$, so $P \times_G Q$ no longer
carries any remnants of the action of~$G$ on~$Q$.}
the only way to formulate the condition that $V$ be globally invariant
is to require $V$ to be invariant under the action of $P \times_G G$.
For a truly dynamical theory, with configuration bundle $P \times_G Q$,
we must of course include derivatives of fields that are sections of~%
$P \times_G Q$. For simplicity, we shall assume, as always in this paper,
that the dynamics of the theory is governed by a first order ``matter
field lagrangian'', for which we contemplate three slightly different
options regarding the choice of its domain,%
\footnote{For the range, we use volume forms (pseudoscalars) rather
than scalars: this allows us to integrate lagrangians over regions of
space-time without having to fix a separately defined volume form on~$M$.}
depending on whether we use the linearized first order jet bundle $\vec{J}%
(P \times_G Q)$ or the full first order jet bundle $J(P \times_G Q)$ of~%
$P \times_G Q$ and whether we include an explicit dependence on connections
or not:

\pagebreak

\begin{itemize}
 \item $\vec{\mathcal{L}}_{\mathrm{mat}}: \vec{J}(P \times_G Q)
       \longrightarrow \bwedge^{\!n\,} T^*M \,$: such a lagrangian
       will be called \textbf{globally invariant} if it is invariant
       under the action (\ref{eq:ACTPGLJPQ}) of the Lie group bundle
       $P \times_G G$ on~$\vec{J}(P \times_G Q)$.
 \item $\mathcal{L}_{\mathrm{mat}}: J(P \times_G Q) \longrightarrow
       \bwedge^{\!n\,} T^*M \,$: such a lagrangian will be called
       \textbf{locally invariant} or \textbf{gauge invariant} if,
       for every compact subset $K$ of space-time~$M$, the action
       functional $\, (\mathcal{S}_{\mathrm{mat}})_K^{}: \Gamma%
       (P \times_G Q) \longrightarrow \mathbb{R} \,$ defined by
       integration of~$\mathcal{L}_{\mathrm{mat}}$ over~$K$, i.e.,
       \begin{equation} \label{eq:ACTMF1}
        (\mathcal{S}_{\mathrm{mat}})_K^{}[\varphi]~
        =~\int_K \mathcal{L}_{\mathrm{mat}}(\varphi,\partial\varphi)~,
       \end{equation}
       is invariant under the action of the group $\mathrm{Gau}(P)$.
 \item $\mathcal{L}_{\mathrm{mat}}: CP \times_M J(P \times_G Q)
       \longrightarrow \bwedge^{\!n\,} T^*M \,$: again, such a
       lagrangian will be called \textbf{locally invariant}
       or \textbf{gauge invariant} if, for every compact
       subset $K$ of space-time~$M$, the action functional
       $\, (\mathcal{S}_{\mathrm{mat}})_K^{}: \Gamma(CP \times_M
       (P \times_G Q)) \longrightarrow \mathbb{R} \,$ defined by
       integration of~$\mathcal{L}_{\mathrm{mat}}$ over~$K$, i.e.,
       \begin{equation} \label{eq:ACTMF2}
        (\mathcal{S}_{\mathrm{mat}})_K^{}[A,\varphi]~
        =~\int_K \mathcal{L}_{\mathrm{mat}}(A,\varphi,\partial\varphi)~,
       \end{equation}
       is invariant under the action of the group $\mathrm{Gau}(P)$.
\end{itemize}
Regarding gauge invariance, we then have the following
\begin{thm} \label{teo:GINV1}~
 The action functional\/ $\mathcal{S}_{\mathrm{mat}}$ defined by
 integration of the lagrangian\/ $\mathcal{L}_{\mathrm{mat}}$ over
 compact subsets of space-time is gauge invariant if and only if
 the lagrangian\/ $\mathcal{L}_{\mathrm{mat}}$ is invariant under
 the action (\ref{eq:ACTJPGJPQ}) of the Lie group bundle\/
 $J(P \times_G G)$ on\/~$J(P \times_G Q)$, in the first case,
 and under the action of the Lie group bundle\/ $J(P \times_G G)$
 on\/~$\, CP \times_M J(P \times_G Q)$ \linebreak that results
 from combining its actions (\ref{eq:ACTJPGCP}) on\/~$CP$ and~%
 (\ref{eq:ACTJPGJPQ}) on\/~$J(P \times_G Q)$, in the second case.
\end{thm}
\proof
 As observed in the preceding sections, the action (\ref{eq:ACTPGPQ})
 of~$P \times_G G$ on~$P \times_G Q$, when lifted to sections, induces
 the standard action of strict automorphisms of~$P$ on sections $\varphi$
 of~$P \times_G Q$, the action (\ref{eq:ACTJPGJPQ}) of~$J(P \times_G G)$
 on~$J(P \times_G Q)$, when lifted to sections, induces the standard action
 of strict automorphisms of~$P$ on sections of~$P \times_G Q$ together with
 their first order derivatives, and finally the action (\ref{eq:ACTJPGCP})
 of~$J(P \times_G G)$ on~$CP$, when lifted to sections, induces the standard
 action of strict automorphisms of~$P$ on connection forms on~$P$ by pull-back.
 Now since gauge transformations do not move the points of space-time,
 invariance of the integral $(\mathcal{S}_{\mathrm{mat}})_K^{}$ over any
 compact subset $K$ of~$M$ is equivalent to invariance of the integrand~%
 $\mathcal{L}_{\mathrm{mat}}$.
\qed

\noindent
It should be noted at this point that without specifying further data,
only the third of the above versions is physically meaningful. Indeed, the
function $\vec{\mathcal{L}}_{\mathrm{mat}}$ of the first version, by itself,
is not an acceptable lagrangian because lagrangians in field theory must be
defined on the first order jet bundle and not on the linearized first order
jet bundle. On the other hand, lagrangians $\mathcal{L}_{\mathrm{mat}}$ as in
the second version, depending only on the matter fields and their first order
partial derivatives, are impossible to construct directly: all known examples
require additional data. (In particular, this holds when the fields are
represented by sections of bundles which may be non-trivial or at least
are not manifestly trivialized.) The only way to overcome these problems,
all in one single stroke, is to introduce some connection in~$P$: this
may either be a fixed (preferrably, flat) background connection which
allows us to identify $\vec{J}(P \times_G Q)$ and $J(P \times_G Q)$
(and is usually introduced tacitly, without ever being mentioned
explicitly), or it may itself be a dynamical variable, as indicated
in the third version above. In the next section, we shall show that
when this is done, there is a natural prescription, called ``minimal
coupling'', that allows us to pass from a globally invariant lagrangian
$\vec{\mathcal{L}}_{\mathrm{mat}}$ on~$\vec{J}(P \times_G Q)$ to a locally
invariant lagrangian $\mathcal{L}_{\mathrm{mat}}$ on~$\, CP \times_M
J(P \times_G Q)$, and back.

In the gauge field sector, the situation is completely analogous. Now the
configuration bundle is just $CP$, and the dynamics of the theory is assumed
to be governed by a first order ``gauge field lagrangian'', which is a map
$\, \mathcal{L}_{\mathrm{gauge}}: J(CP) \longrightarrow \bwedge^{\!n\,} T^*M \,$.
(The standard example is of course the Yang-Mills lagrangian.) Again, such
a lagrangian will be called \textbf{locally invariant} or \textbf{gauge
invariant} if, for every compact subset $K$ of space-time~$M$, the action
functional $\, (\mathcal{S}_{\mathrm{gauge}})_K^{}: \Gamma(CP) \longrightarrow
\mathbb{R} \,$ defined by integration of $\mathcal{L}_{\mathrm{gauge}}$ over~$K$,
i.e.,
\begin{equation} \label{eq:ACTGF}
 (\mathcal{S}_{\mathrm{gauge}})_K^{}[A]~
 =~\int_K \mathcal{L}_{\mathrm{gauge}}(A,\partial A)~,
\end{equation}
is invariant under the action of the group $\mathrm{Gau}(P)$.
\begin{thm} \label{teo:GINV2}~
 The action functional\/ $\mathcal{S}_{\mathrm{gauge}}$ defined by integration
 of the lagrangian\/ $\mathcal{L}_{\mathrm{gauge}}$ over compact subsets
 of space-time is gauge invariant if and only if the lagrangian\/
 $\mathcal{L}_{\mathrm{gauge}}$ is invariant under the action~%
 (\ref{eq:ACTJ2PGJCP}) of the Lie group bundle\/ $J^2(P\times_G G)$
 on\/~$J(CP)$.
\end{thm}
\proof
 As observed in the preceding sections, the action (\ref{eq:ACTJPGCP})
 of~$J(P \times_G G)$ on~$CP$, when lifted to sections, induces the
 standard action of strict automorphisms of~$P$ on connection forms
 $A$ on~$P$ and the action (\ref{eq:ACTJ2PGJCP}) of~$J^2(P \times_G G)$
 on~$J(CP)$, when lifted to sections, induces the standard action of
 strict automorphisms of~$P$ on connection forms on~$P$ together with
 their first order derivatives. Now since gauge transformations
 do not move the points of space-time, invariance of the integral
 $(\mathcal{S}_{\mathrm{mat}})_K^{}$ over any compact subset $K$ of~$M$
 is equivalent to invariance of the integrand~$\mathcal{L}_{\mathrm{mat}}$.
\qed

The general situation is handled by simply combining the previous
constructions. The configuration bundle is now $\, E = CP \times_M
(P \times_G Q)$, as stated at the beginning of the section, and the
dynamics of the complete theory is governed by a total lagrangian
$\, \mathcal{L}: JE \longrightarrow \bwedge^{\!n\,} T^*M \,$ which is
the sum of a ``gauge field lagrangian'' $\mathcal{L}_{\mathrm{gauge}}$,
as before, and a ``matter field lagrangian'' $\, \mathcal{L}_{\mathrm{mat}}:
CP \times_M J(P \times_G Q) \longrightarrow \bwedge^{\!n\,} T^*M \,$:
\begin{equation} \label{eq:LAGTOT}
 \mathcal{L}(A,\partial A,\varphi,\partial\varphi)~
 =~\mathcal{L}_{\mathrm{gauge}}(A,\partial A) \, + \,
   \mathcal{L}_{\mathrm{mat}}(A,\varphi,\partial\varphi)~.
\end{equation}
(Note that the gauge fields do appear in the matter field lagrangian~--
otherwise, there would be no coupling between matter fields and gauge
fields and hence no interaction~-- but appear only as auxiliary fields,
or Lagrange multipliers, i.e., through the gauge potentials without
any derivatives.) The hypothesis of local (or gauge) invariance is
then understood to mean that both $\mathcal{L}_{\mathrm{gauge}}$ and
$\mathcal{L}_{\mathrm{mat}}$ should be locally (or gauge) invariant,
and so Theorems~\ref{teo:GINV1} and~\ref{teo:GINV2} apply as before.

An even stronger hypothesis than gauge invariance, which might be called gauge
plus space-time diffeomorphism invariance, would be that these functionals
are invariant under the action of the whole group $\mathrm{Aut}(P)$.
However, this condition is really much too strong: \linebreak
it is realized to some extent in ``topological field theories'', the standard
examples of which are Chern-Simons theories in~$3$-dimensional space-time,
but it certainly does not hold for gauge theories that are of direct
physical relevance. What might be considered is invariance under sub%
groups of $\mathrm{Aut}(P)$ that cover space-time isometries (Poincar\'e
transformations, in the case of flat Minkowski space-time), rather than
arbitrary diffeomorphisms.

\section{Minimal Coupling}

In this section, we want to show that the prescription of ``minimal coupling'',
which plays a central role in the construction of gauge invariant lagrangians
for the matter field sector, can be understood mathematically as a one-to-one
correspondence between globally invariant matter field lagrangians
\begin{equation}
 \vec{\mathcal{L}}_{\mathrm{mat}}: \vec{J}(P \times_G Q)~
 \longrightarrow~\bwedge^{\!n\,} T^*M~,
\end{equation}
and locally invariant matter field lagrangians
\begin{equation}
 \mathcal{L}_{\mathrm{mat}}: CP \times_M J(P \times_G Q)~
 \longrightarrow~\bwedge^{\!n\,} T^*M~.
\end{equation}
The basic idea underlying this correspondence can be summarized
in a simple commutative diagram,
\begin{equation} \label{eq:ACOMIN}
 \begin{array}{c}
  \xymatrix{
   ~J(P \times_G G)~ \ar[dd] \ar@{-->}[rr]^-{\mbox{acts}}_-{\mbox{on}} & &
   ~CP \times_M J(P \times_G Q)~ \ar[dd]_-{D}
     \ar[dr]_-{\mathcal{L}_{\mathrm{mat}}} & \\
   & & & \mbox{}~~~\bwedge^{\!n\,} T^* M~, \\
   ~P \times_G G~ \ar@{-->}[rr]^-{\mbox{acts}}_-{\mbox{on}} & &
   ~\vec{J}(P \times_G Q)~ \ar[ur]^-{\vec{\mathcal{L}}_{\mathrm{mat}}} &
  }
 \end{array}
\end{equation}
where the first vertical arrow is the target projection from $J(P \times_G G)$
to $P \times_G G$ while the second vertical arrow is the \textbf{covariant
derivative map} defined by
\begin{equation}
 D(A,\varphi,\partial\varphi)~=~(\varphi,D_{\!A} \varphi)~, 
\end{equation}
where, for any local section $\varphi$ of~$P \times_G Q$, its covariant
derivative $D_{\!A} \varphi$ with respect to $A$ can be defined simply as
the composition of its standard derivative $\partial\varphi$ with the
corresponding vertical projection. As indicated in the diagram, this
map is equivariant under the respective actions of the Lie group
bundles $J(P \times_G G)$ and $P \times_G G$, and even more than
that is true: it takes the $J(P \times_G G)$-orbits in $\, CP \times_M
J(P \times_G Q) \,$ precisely onto the $(P \times_G G)$-orbits in~%
$\vec{J}(P \times_G Q)$. This proves our claim that the formula
\begin{equation}
 \mathcal{L}_{\mathrm{mat}}(A,\varphi,\partial\varphi)~
 =~\vec{\mathcal{L}}_{\mathrm{mat}}(\varphi,D_A \varphi)~.
\end{equation}
establishes a one-to-one correspondence between ($\bwedge^{\!n\,}
T^* M$)-valued functions $\mathcal{L}_{\mathrm{mat}}$ on $\, CP \times_M
J(P \times_G Q) \,$ and ($\bwedge^{\!n\,} T^* M$)-valued functions
$\vec{\mathcal{L}}_{\mathrm{mat}}$ on $\vec{J}(P \times_G Q)$: this
construction of the former from the latter is what is known as the
prescription of \textbf{minimal coupling}.

\section{Utiyama's Theorem}

Another important fact concerning the construction of gauge invariant
lagrangians, this time in the gauge field sector, is known as Utiyama's
theorem: it states, roughly speaking,  that any gauge invariant lagrangian
in the gauge field sector must be a function only of the curvature tensor
and its covariant derivatives. In the present context where we consider
only first order lagrangians, it can be understood mathematically as a
one-to-one correspondence between globally invariant lagrangians
\begin{equation}
 \mathcal{L}_{\mathrm{curv}}:
 \bwedge^{\!2\,} T^* M \otimes (P \times_G \mathfrak{g})~
 \longrightarrow~\bwedge^{\!n\,} T^*M~,
\end{equation}
and locally invariant gauge field lagrangians
\begin{equation}
 \mathcal{L}_{\mathrm{gauge}}: J(CP)~\longrightarrow~\bwedge^{\!n\,} T^*M~.
\end{equation}
Again, the basic idea underlying this correspondence can be summarized
in a simple commutative diagram
\begin{equation} \label{eq:UTIYAM}
 \begin{array}{c}
  \xymatrix{
   ~J^2(P \times_G G)~ \ar[dd] \ar@{-->}[rr]^-{\mbox{acts}}_-{\mbox{on}} & &
   ~J(CP)~ \ar[dd]_-{F}
   \ar[dr]_-{\mathcal{L}_{\mathrm{gauge}}} & \\
   & & & \mbox{}~~~\bwedge^{\!n\,} T^* M~, \\
   ~P \times_G G~ \ar@{-->}[rr]^-{\mbox{acts}}_-{\mbox{on}} & &
   ~\bwedge^{\!2\,} T^* M \otimes (P \times_G \mathfrak{g})~
   \ar[ur]^-{\mathcal{L}_{\mathrm{curv}}} &
  }
 \end{array}
\end{equation}
where the first vertical arrow is the target projection from
$J^2(P \times_G G)$ to $P \times_G G$ while the second vertical
arrow is the \textbf{curvature map} that to each connection form
$A$ associates its curvature form $F_A$. As indicated in the diagram,
this map is equivariant under the respective actions of the Lie group
bundles $J^2(P \times_G G)$ and $P \times_G G$, and as stated in
Proposition~\ref{prp:ACTJ2PGJCP}, it takes the $J^2(P \times_G G)$-%
orbits in~$J(CP)$ precisely onto the $(P \times_G G)$-orbits in~%
$\, \bwedge^{\!2\,} T^* M \otimes (P \times_G \mathfrak{g})$.
This proves our claim that the formula
\begin{equation}
 \mathcal{L}_{\mathrm{gauge}}(A,\partial A)~
 =~\mathcal{L}_{\mathrm{curv}}(F_A)~.
\end{equation}
establishes a one-to-one correspondence between ($\bwedge^{\!n\,}
T^* M$)-valued functions $\mathcal{L}_{\mathrm{gauge}}$ on $J(CP)$ and
($\bwedge^{\!n\,} T^* M$)-valued functions $\mathcal{L}_{\mathrm{curv}}$
on~$\bwedge^{\!2\,} T^* M \otimes (P \times_G \mathfrak{g})$.

\section{Conclusions and Outlook}

As we have tried to demonstrate in this paper, the appropriate mathematical
concept for dealing with symmetries in classical field theory, when adopting
a geometrical and at the same time purely finite-dimensional framework (as
opposed to a functional approach), is that of Lie group bundles and their
actions on fiber bundles over space-time~$M$ (whose sections constitute the
fields of the model at hand). This general statement applies to internal
symmetries, global as well as local, allowing to view the passage from the
former to the latter~-- generally known as the procedure of ``gauging a
symmetry''~-- simply as the transition from the original Lie group bundle
to its jet bundle. It also allows for a conceptually transparent and natural
formulation of various procedures and statements that play an important role
in gauge theories, such as the prescription of ``minimal coupling'' \linebreak
and Utiyama's theorem on the possible form of gauge invariant lagrangians for
the pure gauge field sector. The entire approach is a generalization of a
corresponding appraoch to classical mechanics~\cite{FK}, to which it reduces
when one takes $\, M = \mathbb{R}$, which implies that the principal $G$-%
bundle $P$ over~$M$ is trivial (and hence so are all other bundles involved),
and when one supposes that a fixed trivialization has been chosen: then the
product in $\, P \times_G G \cong \mathbb{R} \times G \,$ and its action
on~$\, P \times_G Q \cong \mathbb{R} \times Q \,$ do not depend on the
base point, or in other words, they reduce to an ordinary Lie group
product in $G$ and an ordinary action of~$G$ on the manifold~$Q$.
Similarly, the induced product in $\, J(P \times_G G) \cong \mathbb{R}
\times TG$ \linebreak and its induced action on $\, J(P \times_G Q) \cong
\mathbb{R} \times TQ \,$ also do not depend on the base point and reduce
to the ordinary induced Lie group product in~$TG$ and the ordinary induced
action of~$TG$ on~$TQ$. These reductions explain why in the case of mechanics,
the need for using Lie group bundles, rather than ordinary Lie groups, was not
properly appreciated.

Finally, the extension of this approach to achieve unification of internal
symmetries with space-time symmetries requires the transition from Lie group
bundles to Lie groupoids~-- a problem which is presently under investigation.
Another issue is how to correctly formulate invariance of geometric objects
represented by certain prescribed tensor fields (such as pseudo-riemannian
metrics or symplectic forms, for instance) under actions of Lie group bundles
or, more generally, Lie groupoids and/or their infinitesimal counterparts,
that is, Lie algebra bundles or, more generally, Lie algebroids. These and
similar questions will have to be answered before one can hope to really
understand what is the field theoretical analogue of the momentum map of
classical mechanics.

\section*{Acknowledgements}

This work has been financially supported by CAPES (``Coordena\c{c}\~ao
de Aperfei\c{c}oamento de Pessoal de N\'{\i}vel Superior''), by CNPq
(``Conselho Nacional de Desenvolvimento Cient\'{\i}fico e Tecnol\'ogico'')
and by FAPESP (``Funda\c{c}\~ao de Amparo \`a Pesquisa do \linebreak
Estado de S\~ao Paulo''), Brazil.

\end{document}